%% file: main.tex
\AddToHook{package/hyperxmp/before}{\RequirePackage[bookmarksnumbered,unicode]{hyperref}}
\documentclass[sigconf,nonacm,dvipsnames,prologue,balance=false]{acmart}

\setcopyright{none}
\usepackage{listings}
\usepackage{lstautogobble}
\usepackage{graphicx}
\usepackage{subcaption}
\usepackage{booktabs}
\usepackage{balance}
\usepackage{tikz}
\usepackage{textcomp}
\usepackage{stfloats}
\usepackage{enumitem}
\setlist{nolistsep}

\newcommand*\filled[1]{\tikz[baseline=(char.base)]{
  \node[shape=circle,fill=Black,inner sep=0.4pt] (char) {\textcolor{white}{#1}};}}

\definecolor{greencomments}{rgb}{0,0.5,0}
\definecolor{redstrings}{rgb}{0.9,0,0}
\definecolor{graynumbers}{rgb}{0.5,0.5,0.5}
\definecolor{background_ok}{rgb}{0.95,0.95,0.95}
\makeatletter
\let\hyxmp@parse@acmart\relax
\makeatother

\newcommand{\setpaperpdfmetadata}{\hypersetup{
  keeppdfinfo,
  pdftitle={Multi-GPU Communication Schemes for Iterative Solvers: When CPUs are Not in Charge},
  pdfauthor={Ismayil Ismayilov, Javid Baydamirli, Doǧan Sağbili, Mohamed Wahib, Didem Unat},
  pdfsubject={This paper proposes a fully autonomous execution model for multi-GPU applications that completely excludes the involvement of the CPU beyond the initial kernel launch.},
  pdfkeywords={Multi-GPU, GPU-initiated communication, Persistent kernels, Iterative solvers, NVSHMEM},
  pdfcreator={LaTeX},
  pdflang={en},
  pdfmetalang={en},
  pdfcontactemail={iismayilov21@ku.edu.tr},
  pdfcontactcountry={Turkey}
}}
\setpaperpdfmetadata

\begin{document}
\title{Multi-GPU Communication Schemes for Iterative Solvers: When CPUs are Not in Charge}

\author{Ismayil Ismayilov}
\email{iismayilov21@ku.edu.tr}
\affiliation{
  \institution{Koç University}
  \country{Turkey}
}
\orcid{0000-0003-1831-5556}

\author{Javid Baydamirli}
\email{jbaydamirli21@ku.edu.tr }
\affiliation{
  \institution{Koç University}
  \country{Turkey}
}
\orcid{0000-0001-7235-6418}

\author{Doǧan Sağbili}
\email{dsagbili17@ku.edu.tr}
\affiliation{
  \institution{Koç University}
  \country{Turkey}
}
\orcid{0000-0002-9603-2466}

\author{Mohamed Wahib}
\email{mohamed.attia@riken.jp}
\affiliation{
  \institution{RIKEN}
  \country{Japan}
}
\orcid{0000-0002-7165-2095}

\author{Didem Unat}
\email{dunat@ku.edu.tr}
\affiliation{
  \institution{Koç University}
  \country{Turkey}
}
\orcid{0000-0002-2351-0770}

\begin{abstract}
This paper proposes a fully autonomous execution model for multi-GPU applications that completely excludes the involvement of the CPU beyond the initial kernel launch.
In a typical multi-GPU application, the host serves as the orchestrator of execution by directly launching kernels, issuing communication calls, and acting as a synchronizer for devices.
We argue that this orchestration, or control flow path, causes undue overhead and can be delegated entirely to devices to improve performance in applications that require communication among peers.
For the proposed {\em CPU-free} execution model, we leverage existing techniques such as persistent kernels, thread block specialization, device-side barriers, and device-initiated communication routines to write fully autonomous multi-GPU code and achieve significantly reduced communication overheads.
We demonstrate our proposed model on two broadly used iterative solvers, 2D/3D Jacobi stencil and Conjugate Gradient(CG).
Compared to the CPU-controlled baselines, the CPU-free model can improve 3D stencil communication latency by 58.8\% and provide a 1.63x speedup for CG on 8 NVIDIA A100 GPUs.
The project code is available at \url{https://github.com/ParCoreLab/CPU-Free-model}.
\end{abstract}

\maketitle
\setpaperpdfmetadata
\pagestyle{plain}

\input{sections/intro}
\input{sections/motivation_new_new.tex}

\input{sections/designs}

\input{sections/implementation}

\input{sections/results}
\input{sections/related-work}

\section{Conclusion}
We propose a fully federated multi-GPU execution model and show its viability in two widely used iterative solvers: Stencil and Conjugate Gradient using persistent kernels, thread block specialization, device-initiated communication, and synchronization.
By eliminating costly host-side communication and synchronization routines and moving the control path to devices completely, we can achieve significantly reduced communication latencies, allowing better overlap when execution time is bound by data movement across devices.
Moreover, we can take advantage of ephemeral on-chip memory across iterations in persistent kernels.
Our experiments on 8 NVIDIA A100 GPUs showed that the CPU-free model could significantly improve performance, especially in communication latency-bounded scenarios.

We view this work as an important next step toward the ultimate goal of complete GPU autonomy.
In line with this, we plan to apply the CPU-free model to more applications and codify precisely when it is advantageous.
We also plan to explore ways to automate porting existing CPU-controlled applications to the CPU-free model.

\section*{Acknowledgments}
This project has received funding from the European Research Council (ERC) under the European Union’s Horizon 2020 research and innovation programme (grant agreement No 949587).

\balance
\input{main.bbl}

\end{document}

%% file: sections/intro.tex
\section{Introduction}

GPUs have become the leading accelerator in modern HPC systems, equipping 7 of the 10 leading Top500 supercomputers in the world \cite{top500}. As real-life applications rely on multiple GPUs to solve large problems, communication among GPUs can quickly become a performance bottleneck, leading to poor application scaling \cite{Peta-scale-stencil}. 
In the traditional model of execution, communication among GPUs in large systems has been mediated through the host.
In the absence of direct peer-to-peer communication, GPU-to-GPU data transfers had to be routed through the CPU, which incurred larger communication latencies \cite{Agostini_Offloading_Communication_Control_Logic, gpudirectrdma}. 
Advances, such as NVLink \cite{nvlink} and GPUDirect RDMA \cite{gpudirectrdma}, have since allowed CUDA devices to bypass the intermediate CPU buffers during data transfers and directly communicate with their peers. 

While data movement can now be delegated to the GPU, the communication control flow still sits firmly on the CPU. Even with direct GPU-to-GPU communication, data transfers are mostly 
\textit{initiated} from the CPU using host-side API calls, putting it in charge of the overall flow of multi-GPU execution; the CPU enqueues both computational kernels and communication calls to devices, overlaps them if possible, and synchronizes them for functional correctness. When required, the CPU also acts as a global barrier for synchronizing multiple devices.

Regardless of whether the communication happens directly between devices, the presence of the CPU in the control path is implicitly assumed. 

This paper argues that freeing the GPU from the host by moving the control flow to devices has several benefits for multi-GPU applications, particularly in communication-bound settings. To illustrate these 
advantages, we propose a distinct \emph{CPU-Free} execution model that removes the CPU from said control path and gives autonomy to GPUs to control their communication and synchronization. In order to realize this, we leverage and combine four essential programming concepts. The first technique  is to switch from kernels launched sequentially by the host to a long-running persistent kernel \cite{persistent-threads, whippletree}. Second, we make use of thread block specialization within our persistent kernel to explicitly overlap communication and computation by reserving a number of blocks for each phase. Third, we adopt GPU-\emph{initiated} communication methods to stage data movement independently of the CPU. Finally, we hand over the task of synchronizing multiple GPUs to the devices themselves and allow them to independently synchronize with their peers.

These changes make GPUs less dependent on the host, allow for more asynchrony, reduce communication latencies, and lead to better communication/computation overlap.

We demonstrate the effectiveness of our execution model on two well-known iterative solvers: 2D/3D Jacobi stencils \cite{stencil} and a Krylov subspace method, namely Conjugate Gradient (CG) \cite{krylov}. The \textit{CPU-free} execution model is a good fit for these types of applications as their iterative nature requires communication and computation at each iteration - tasks otherwise managed by the CPU. 

\begin{figure*}[th]
\centering
\begin{subfigure}{0.32\textwidth}
\input{listings/cpu-controlled.tex}
\caption{\centering \footnotesize{Pseudo-code of CPU-controlled stencil solver \cite{nvidia-multi-gpu-programming-models}}}
\label{fig:cpu-controlled-code}
\end{subfigure}
\hspace{2pt}
\begin{subfigure}{0.32\textwidth}
\includegraphics[width=\textwidth]{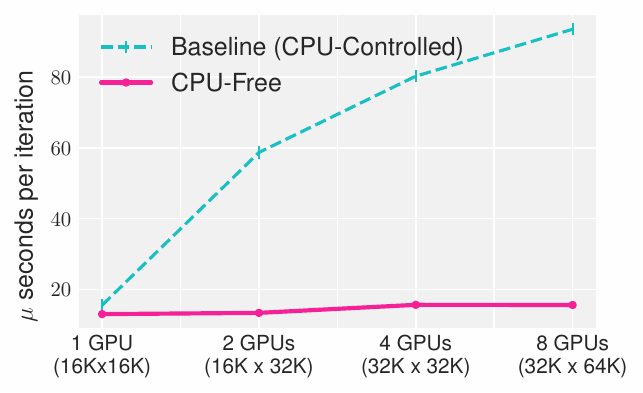}
\caption{\centering \footnotesize{Communication overhead with no computation}}
\label{fig:nocomp}
\end{subfigure}
\hspace{2pt}
\begin{subfigure}{0.32\textwidth}
\includegraphics[width=\textwidth]{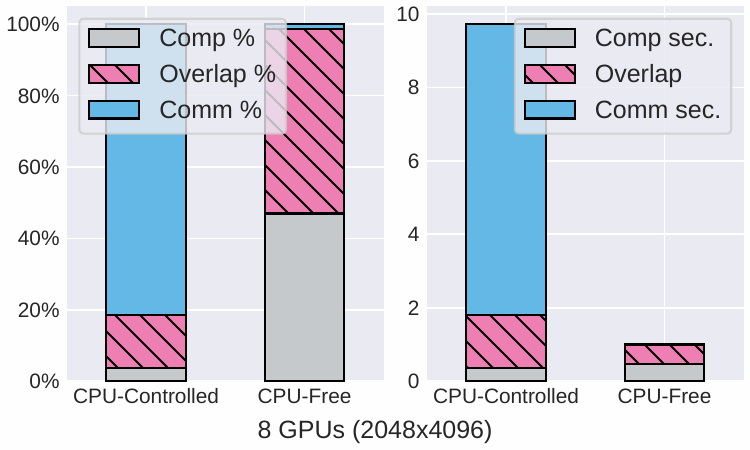}
 \caption{\centering \footnotesize{Communication overlap percentage for 2D stencil}}
\label{fig:percentage-overlap-medium}
\end{subfigure}
\caption{(a) Pseudo-code of CPU-controlled 2D stencil example serving as a baseline for comparison 
(b) Communication and synch overheads with no computation of our CPU-free model against CPU-controlled 
(c) Left shows communication overlap percentages for CPU-free and CPU-controlled models. Right shows the execution times.}
\label{fig:motivation}
\end{figure*}

Overall, our paper makes the following contributions:
\begin{itemize}[align=left,leftmargin=*,widest={10}]
    \item We describe a CPU-free execution model that 
    removes the host from the critical path, grants autonomy to devices, reduces latencies, and achieves better communication/computation overlap. 
    \item We design and implement iterative Jacobi 2D/3D and Conjugate Gradient solvers on multiple GPUs using our CPU-free execution model. 
    \item We evaluate the CPU-free execution model and compare its performance against CPU-controlled baselines using 8 NVIDIA A100 GPUs. Stencil kernels enjoy anywhere from $18.8$\% to $96.2$\% improvement in performance, depending on the domain size, by significantly lowering communication latencies and allowing better overlap.
    Standard CG and Pipelined CG variants observe 1.54x and 1.63x speedup on 18 sparse matrices compared to their CPU-controlled counterparts.
 
\end{itemize}

As noted, the CPU-free model's individual components have been previously proposed in other contexts. For instance, earlier works have shown the benefits of employing persistent kernels to reduce CPU involvement.
A recent work by Zhang et al. \cite{perks} utilizes the GPU's shared memory and register volumes across iterations to avoid costly trips to global memory in persistent kernels.
The idea of reducing CPU involvement has been studied in the literature to accelerate graph algorithms \cite{atos, atos-single, nvshmem-3}.
Other studies have proposed different methods for optimizing GPU networking by directly initiating communication from the device \cite{Agostini_Offloading_Communication_Control_Logic, dCUDA, gpudirectrdma, ComP-Net, lena-oden, nvshmem-2, gravel}, as well as inter-kernel networking operations and solutions to related correctness issues \cite{GIO}.
However, no prior art combines these components to create a systematic way of implementing CPU-free applications.
The paper serves as a practical proof of concept showing one can get significant performance improvements, especially for communication latency bounded cases, by completely cutting out the CPU and provides a blueprint for how to do this for broadly used iterative solvers.

%% file: listings/cpu-controlled.tex
\begin{lstlisting}[backgroundcolor=\color{background_ok}, basicstyle=\ttfamily\tiny,label={lst:cpu-controlled}]
    // |\filled{1}| Time loop on CPU
    while (iter++ < num_iterations) {     
        // |\filled{2}|  Launch compute kernel
        stencil_kernel<<<..., comp_stream>>>(...)  
        // |\filled{3}|  Sync with neighboring GPUs
        wait_neighbors(comm_stream)
        // |\filled{4}|  Compute and communicate boundaries
        compute_boundaries<<<..., comm_stream>>>(...)
        write_neighbors(comm_stream)
        // |\filled{5}|  Sync comm and comp streams
        sync(comm_stream, comp_stream)
    }
\end{lstlisting}

%% file: sections/motivation_new_new.tex
\section{Background \& Motivation}

In this work, we seek to give devices full autonomy by decoupling them from the CPU during multi-GPU execution. We first note that by the advent of GPU-initiated communication \cite{extendingOpenShmem, gpudirectrdma, nvshmem-2}, the {\em data path} - the path in which data is transferred through communication routines - can be moved entirely to the GPU. CUDA devices can perform data transfers among one another through either NVLink\cite{nvlink} or PCIe within a node or through NICs in multi-node systems, both avoiding intermediate host-side transfers.

Despite the progress made in pushing the data path to the GPU, overall control of execution - the {\em control path} -  has stayed resident on the host. First, in single-GPU applications, the CPU serves as a barrier that synchronizes kernels with each other. This synchronization is necessary 

in the case of any iterative solver; to ensure correctness, the computations for timestep $T+1$ can start only after timestep $T$ concludes. The CPU satisfies this requirement by launching a kernel every iteration since kernels launched back-to-back are guaranteed to execute serially by the GPU scheduler. This way, the kernel launch and teardown act as an \textit{implicit} barrier between timesteps. 
For the rest of this work, we refer to such implicitly-synchronized kernels as \textit{discrete} kernels.

For multi-GPU applications, the CPU plays a similarly active role. In addition to kernel launches, the CPU also issues the communication using host-side APIs (i.e., `{\tt cudaMemcpy}`). These calls are issued on the CPU even when the underlying transmission uses the direct GPU-to-GPU data path. When applications allow for overlap between independent phases of communication and computation, it is again the CPU that orchestrates this overlap by enqueueing communication and computation on separate GPU streams 

to run concurrently and then synchronize through GPU events. Moreover, the CPU acts as a global barrier when synchronizing multiple GPUs, possibly using CPU-side barriers.

Listing \ref{fig:cpu-controlled-code} illustrates the CPU's involvement in the control path with a stencil solver using the \textit{CPU-controlled} model of execution \footnote{Based on multi-threaded copy overlap \href{https://github.com/NVIDIA/multi-gpu-programming-models}{as provided by NVIDIA}}. 
\filled{1} First, the CPU maintains a time loop that invokes the computational kernel, communication, and synchronization calls every iteration. \filled{2} The CPU launches the compute kernel in a {\tt comp\_stream}, which performs the stencil operations. \filled{3} The CPU synchronizes with neighbors to ensure that inbound halos have been received. \filled{4} Next, the CPU launches a kernel to compute the boundary rows and communicates them as the neighbors' halos. Both of these routines are enqueued to a different stream, namely {\tt comm\_stream}, to achieve overlap between communication and computation. 
Thus, step \filled{2} is overlapped with step \filled{3} and \filled{4}. 
\filled{5} Finally, the CPU synchronizes these two streams before advancing to the next iteration.

Figure \ref{fig:motivation}b presents the communication overheads with no computation - both synchronization and data movement across GPUs - of CPU-controlled and CPU-free executions in increasing problem sizes for the 2D Jacobi solver. The CPU-controlled baseline uses the  overlapping implementation shown in Listing \ref{fig:cpu-controlled-code}, utilizing host-side CUDA events and memory copy calls to explicitly overlap communication and computation. The CPU-free version instead uses GPU-initiat\-ed calls for both of those routines. We notice significantly and consistently lower communication overheads in the CPU-free version in all domain sizes - up to 5.9x at 8 GPUs. 

Figure \ref{fig:percentage-overlap-medium} shows the ratio of computation and communication overlap of both versions, along with leftover communication overhead that could not be overlapped. We note the overflown communication time in the baseline, where it takes over $96$\% of the execution, of which only $19$\% is overlapped with computation, leading to suboptimal performance. We reclaim $89.7$\% of the full execution time by reducing the communication latency $17.5$ times. 
Though high communication latency can be hidden easily in large domains when computation time dwarfs it,
CPU-free execution is particularly of high importance in strong scaling cases and also in phases in simulations where the workload drops (for example, computing boundary conditions for 2D planes in a 3D discretized domain to solve PDEs~\cite{DBLP:conf/sc/WahibM14,DBLP:conf/iccS/YamaguchiFIHHW17,DBLP:journals/softx/AfanasyevBMOTVF21}).

%% file: sections/designs.tex
\section{CPU-Free Execution Model}
\label{sec:designs}

\subsection{Overview}
We propose an execution model that eliminates the CPU from both the data and control paths to achieve performance benefits. In order to give full autonomy to GPUs, we make use of several prerequisites and techniques, such as persistent kernels, thread-block specialization, GPU-initiated data transfers, and  device-side synchronization. 

\begin{figure}[t]
     \centering
     \includegraphics[width=0.8\linewidth]{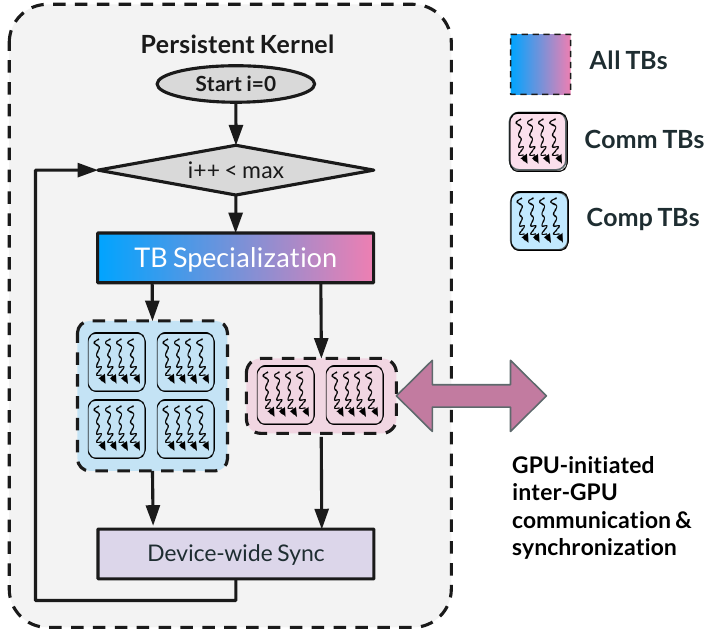}
     \caption{CPU-free execution model leverages persistent kernels, thread-block (TB) specialization, GPU-initiated data transfers, and device-side synchronization. Comm: Communication, Comp: Computation.}
     \label{fig:cpu-free-overview}
 \end{figure}

Figure \ref{fig:cpu-free-overview} shows the concepts that constitute the basis of the CPU-free execution model.

\subsubsection{Persistent Kernels}
Traditionally, iterative GPU kernels are implemented on a per-iteration basis, meaning a new instance of the kernel is enqueued in a stream for each time step. 
The GPU is oblivious to the iterative nature of the computation and supplementary operations. 

This kind of execution relies on the CPU to provide implicit device synchronization across iterations to ensure correctness, as most iterative solvers have temporal dependencies on preceding timesteps. To eliminate such CPU dependence, we implement a long-running persistent kernel, originally proposed in \cite{persistent-threads}, into which we move the time loop of the solver. This allows us to grant more autonomy to the GPU by freeing it from needing to return to the CPU each iteration.

\subsubsection{Thread Block Specialization}
Key in our design is the nascent idea of \textit{thread block specialization} whereby TBs may work on different tasks within the same kernel (conceptually similar to warp specialization~\cite{10.1145/2555243.2555258}). 
 We utilize this idea to achieve communication/computation overlap to reduce communication latency. 

Discrete kernels achieve overlap by enqueueing communication and computation kernels in concurrent streams through host-side runtime calls and transferring data with asynchronous {\tt Memcpy} calls that run independently in copy engines.
We accomplish this asynchronous behavior on the device side by specializing a number of TBs within a kernel to manage communication while the remainder of the TBs handle the bulk of the computation.

\subsubsection{GPU Initiated Data Movement}
We use direct GPU-to-GPU data transfers \emph{within} the kernel and initiate GPU communication among peers without host involvement. Direct GPU-to-GPU data movement is used in two ways: to communicate the actual data and to synchronize neighboring GPUs. We use NVSHMEM \cite{nvshmem-2, Nvshmem_Docs}, NVIDIA's implementation of OpenSHMEM \cite{openshmem}, for all direct GPU-initiated data transfers, as it provides a fine-grained device-side API for data movement.

\subsubsection{Device-Side Synchronization}
Synchronization in a traditional CUDA kernel is limited to threads within a single thread block, and
the kernel launch itself acts as a barrier. Instead, we use device-wide barriers to synchronize the thread blocks across the persistent kernel. At the end of each iteration, device-wide barriers introduced in CUDA 9.0 with the Cooperative Groups API are used to synchronize the communication and computation TBs. Although the latency difference between implicit synchronization using sequential kernel launches and explicit synchronization within the kernel through grid sync is negligible \cite{sync-study}, it is no longer required for the CPU to orchestrate the kernel launches just to synchronize threads within a kernel. 

For synchronization between peer GPUs, traditionally, host-side methods, such as OpenMP and MPI barriers, are used. We move this control back to the GPU using NVSHMEM's device-side signaling operations for peer device synchronization.

\input{sections/benefits.tex}
\input{sections/usecases.tex}

%% file: sections/benefits.tex
\vspace{-1.5mm}
\subsection{Benefits of CPU-free Execution}

As a promising alternative to traditional accelerator programming,  the CPU-free execution model  has the following benefits over CPU-controlled multi-GPU programming. 

\begin{enumerate}[align=left,leftmargin=*,widest={10}]
         \item {\bf Reduced kernel launch/CPU synchronization overheads.} Traditional implementations require multiple API calls to overlap communication and computation. These operations would be launched in separate streams to run concurrently and synchronized through the host. The CPU-free model allows for greater degrees of kernel fusion as the communication and computation can be put inside one fused kernel. This is beneficial because \textbf{(i)} one fat kernel substantially reduces the kernel launch overheads (Figure \ref{fig:cg-strong-scaling}) and \textbf{(ii)} it eliminates the frequent GPU-CPU synchronization overheads, whose latencies can quickly add up \cite{sync-study}. 

     \item {\bf Reduced communication overheads}. We reduce the communication overheads by initiating communication on the device side.
        
        Since communication calls can be issued on the device side within the kernel at any point by the GPU as soon as data is ready, it provides  more asynchrony by allowing the programmer to inline the communication with the computation. In addition, we reduce cross-GPU synchronization latency by using device-side signaling operations (Figure \ref{fig:nocomp}).

      \item  {\bf Communication/Computation overlap}. In tradition\-al implementations, adequate communication / computation overlap can only be achieved when the domain size saturates the device (all threads are busy).
      But as the problem size per GPU decreases, the kernel launch overheads and the API call latencies can start to dominate the runtime. Because of this, little to no overlap is achieved as the GPU spends a significant amount of time idle. In line with prior work, \cite{gpu-key-value}, we observe that persistent kernels are able to achieve good overlap even when the domain size is small (as shown in Figure \ref{fig:percentage-overlap-medium}). 

      \item {\bf Caching.} We extend the lifetime of a kernel by moving the time loop from the host to the device. This enables the use of the large volume of shared memory, registers to cache intermediate results, and prevents wiping out on-chip memory. 
      As demonstrated by the work by \cite{perks}, this is particularly beneficial for iterative kernels with temporal dependencies between iterations, as the cache would otherwise be destroyed at each time step in a discrete kernel. We observe the performance benefits of caching in stencil methods  from maintaining a single kernel throughout the computation (Figure \ref{fig:2D-Weak-Scaling-all}).

\end{enumerate}

%% file: sections/usecases.tex
\input{sections/stencil.tex}
\label{sec:designs:stencil}

\input{sections/CG.tex}

\label{sec:designs:cg}

%% file: sections/stencil.tex
\begin{figure}[!t]
\centering
\includegraphics[width=\linewidth]{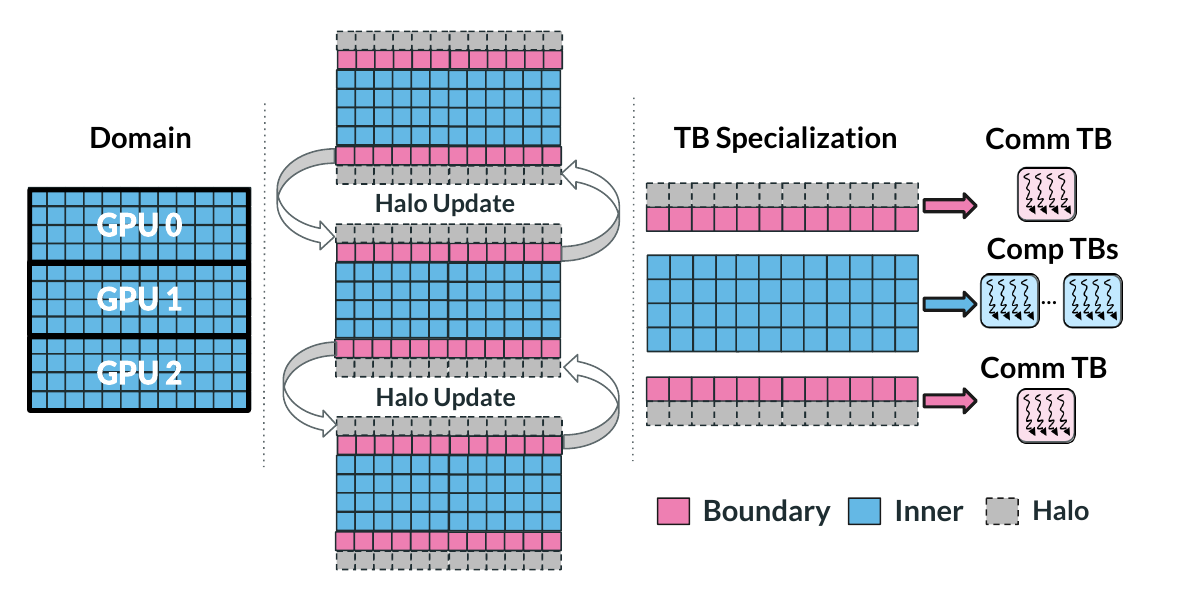}
\caption{Domain decomposition, halo updates, and thread block(TB) specialization for 2D5pt stencil. While communication(comm) TBs handle the boundary computation and halo update with neighbors, computation(comp) TBs handle the inner domain. }
\label{fig:stencil}
\vspace{-5mm}
\end{figure}

\subsection{Use-Case 1: Stencil Methods}
Iterative stencil solvers \cite{strikwerda2004finite} are natural candidates for the proposed execution model as their temporal dependence across the domain requires data movement among all devices and synchronization at each iteration. 

Thus, we present a CPU-free implementation of multi-GPU Jacobi stencils with overlapped communication as our first use case.

The stencil domain can be divided equally among GPUs and split into two independent parts in each device: the inner domain and the boundary, as illustrated in Figure \ref{fig:stencil}. Halo regions that refer to values in neighboring GPUs' domains are appended to each chunk and require communication at each time step. 

As proposed in \cite{Peta-scale-stencil},  we can compute the boundary region independently of the inner domain and communicate concurrently with neighbors while the bulk of the inner domain is being processed. 

\input{listings/stencil_code.tex}

The overlapping design we propose utilizes a persistent kernel that specializes a number of thread blocks to compute the boundary region and issue communication routines as shown in
Listing \ref{lst:stencil-code}. We move the time loop inside the kernel and specialize two concurrent thread blocks for the boundary region (\filled{a}, \filled{b}) and the remainder of the device (\filled{c}) for the inner domain. Each of the boundary thread blocks \filled{1} initially waits on a flag for their corresponding neighbor to signal the availability of the halo region, and \filled{2} uses the said values to compute its boundary region. They then \filled{3} proceed to commit the new values into buffers in their neighbors' memories, and \filled{4} signal them of their availability. \filled{5} Finally, all thread blocks are synchronized at the end of the loop before continuing to the next iteration.

An alternative design to specializing thread blocks in one kernel is to have two co-resident persistent kernels in separate streams, managing boundary and inner domain processing independently. This kind of configuration is more modular and easier to adapt to existing single-GPU kernels but requires an extra sync point between the local stream pairs in each GPU. However, we did not observe any significant performance improvement or degradation from this design compared to the single-stream version.  

 Iterative methods, as discussed earlier, benefit greatly from long-running kernels, as the intermediate results can be cached to be used in subsequent time steps instead of being committed to global memory every iteration. Recent work by Zhang et al. \cite{perks}, namely PERKS, demonstrates this by explicitly caching a portion of the domain in registers and shared memory across iterations through a priority-based caching scheme. We apply our communication scheme on top of the single-GPU stencil implementation of PERKS and extend it with minimal intrusions to the upstream kernel.

%% file: listings/stencil_code.tex
\begin{figure}[t]
\begin{lstlisting}[caption=Stencil kernel using CPU-free  model, label=lst:stencil-code]
|\color{Blue}\_\_global\_\_| void CPU_Free_Jacobi(...) {
    // Time loop on GPU
    while (iter++ < num_iterations) {
        //|\filled{a}| Compute boundary using top neighbor's values
        if (TB_index == 0) {
            //|\filled{1}| Wait for top neighbor to signal
            |\color{Purple}wait\_top\_neighbor|(...)
            //|\filled{2}| Compute top boundary using halos
            top_boundary = compute(top_halo, south, ...)
            //|\filled{3}| Write to top neighbor's bottom halo
            |\color{Purple}write\_top\_neighbor|(top_boundary)
            //|\filled{4}| Signal top neighbor that iteration is done
            |\color{Purple}signal\_top\_neighbor|(...)
        }
        //|\filled{b}| Compute boundary with bottom neighbor's values
        if (TB_index == 1) { ... }           
        //|\filled{c}| Remaining TBs compute the inner domain
        if (TB_index == <rest>) { compute_inner() }
        //|\filled{5}| Synchronize all TBs in kernel
        grid.|\color{Purple}sync|()
}}

\end{lstlisting}
\vspace{-5mm}
\end{figure}

%% file: sections/CG.tex
\begin{figure}[!tbp]
\centering
\includegraphics[width=0.8\linewidth]{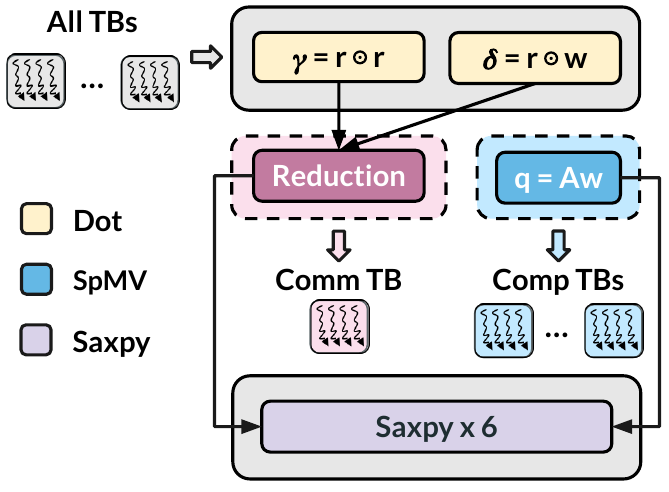}
\caption{Communication/computation overlap with TB specialization for Pipelined CG where reduction of dot products is overlapped with SpMV. Details omitted.  }
\label{fig:CG}
\vspace{-4mm}
 \end{figure}
 
\subsection{Use-Case 2: Conjugate Gradient}
\label{subsec:conjugate-gradient-use-case}
To further demonstrate the broad applicability of our CPU-free execution model, we apply it on top of the Conjugate Gradient (CG) solver, an iterative method used to solve linear equations of the form $Ax = b$ where $A$ is a symmetric and positive definite matrix. 

Three primary operations underlie the CG solver: Sparse matrix-vector multiplication (SpMV), dot product, and  Saxpy.  

Assuming vectors are equally divided among devices when  adapting the method to a multi-GPU setting, Saxpys are vector operations that perform element-wise multiplication and addition on local data and do not require communication.
On the other hand, the computation for SpMV needs vector entries on neighboring GPUs, and the dot product requires a global reduction and a follow-up synchronization over all devices to sum the per-GPU dot contributions.  While both SpMV and dot products involve communication, prior work has shown that the global reduction and synchronization steps for dot products can become a bottleneck when scaling parallel CG to more nodes \cite{cg-synchronization-bottleneck-1, pipelined-CG}.

The commonly used CG algorithm involves Saxpy(x3),  dot product(x2), and SpMV(x1) operations and 

 has few opportunities for communication-computation overlap because of strict dependencies between these operations. To take full advantage of the CPU-free execution model, we also rely on the pipelined CG variant \cite{pipelined-CG, Pipeline-CG-2}, which introduces auxiliary vectors to break up dependencies allowing for the dot product reductions to be overlapped with the computation for SpMV, at the expense of three additional Saxpys. 
Furthermore, because the dot products are now also independent of each other, their communication can be implemented as a single reduction, thus, removing a costly global synchronization step \cite{pipelined-CG}. 
We refer to the traditional and pipelined variants as \textit{Standard CG} and \textit{Pipelined CG} respectively, and implement our design for both variants.  

Figure \ref{fig:CG} illustrates the communication/computation overlap we employ in our Pipelined CG implementation, where the global reduction of dot products is overlapped with SpMV. 

Differently from the stencil case, the \textit{communication} TB does not at all times only communicate; it exclusively handles communication \textit{only} when overlap can be achieved. In parts of execution where no overlap is possible, it joins the \textit{compute} TBs to help with the computation.

Listing \ref{lst:pipelined_cg_pseudocode} shows the pseudocode of the CPU-free  Pipelined CG that uses this overlapping scheme. 

The traditional Pipelined CG implementation would launch up to nine compute kernels every iteration to perform Saxpy(x6), dot products(x2), and SpMV(x1) operations. Our design leverages persistent kernels not only to fuse those nine operations into a single fat kernel but also to move the time loop to the GPU.

In Listing \ref{lst:pipelined_cg_pseudocode}, step \filled{1} shows the time loop running on the device. 
Step \filled{2} leans on the idea of \textit{TB specialization} for communication-computation overlap by reserving one thread block for the global dot reductions (communication) while the remaining TBs handle SpMV (computation).  

Step \filled{3} uses a GPU-initiated reduce call to sum over the local dot contributions. \filled{4} and \filled{5}  remove the CPU's involvement in synchronization by using grid sync to sync \textit{within} the device and a GPU-side global barrier to sync \textit{across} devices.

\input{listings/pipelined_cg_pseudocode.tex}

%% file: listings/pipelined_cg_pseudocode.tex
\begin{figure}[t]
\begin{lstlisting}[caption=Pipelined CG kernel using CPU-free  model, label=lst:pipelined_cg_pseudocode]
|\color{Blue}\_\_global\_\_| void CPU_Free_PipelinedCG(...) {    
    // |\filled{1}| Time loop on GPU
    while (iter++ < num_iterations) {
        |local\_dot|(r, r, gamma)
        |local\_dot|(r, w, delta)
        ...
        // |\filled{2}| Specialize one TB for communication
        if (TB_index == 0) {
            // |\filled{3}| Multi-GPU reduction
            |\color{Purple}sum\_reduce|(gamma, delta) 
        } else {
            |\color{Blue}SpMV|(A, w, q)
        }
        // |\filled{4}| Sync within device
        grid.|\color{Purple}sync|()
        ...
        |saxpy|(...) //x6     
        // |\filled{5}| Sync across devices
        multi_gpu|.\color{Purple}sync|() 
}}
\end{lstlisting}
\vspace{-5mm}
\end{figure}

%% file: sections/implementation.tex
\section{Implementation}

To implement the CPU-free model on NVIDIA GPUs, we utilize the CUDA Cooperative Groups API and NVSHMEM to enable communication and computation entirely within the kernel.
It is worth noting that this approach could also be applied to AMD GPUs since ROCm supports the Cooperative Groups API, and ROC\_SHMEM is functionally equivalent to NVSHMEM \cite{GIO}.
\label{sec:imp}
\subsection{Stencil Methods}
\label{sec:impl:stencil}
We based our stencil code on NVIDIA's open source
\href{https://github.com/NVIDIA/multi-gpu-programming-models}{multi-gpu
    programming models} repository \cite{nvidia-multi-gpu-programming-models}. Though the samples do not necessarily have the most optimized computational kernels - the repository focuses exclusively on
communication optimizations - we refer to it as our main baseline for direct
comparisons of communication overhead. 

\subsubsection{Synchronization and Halo Exchange}

In order to synchronize all thread blocks at the end of a given timestep, we launch our kernels using the CUDA Cooperative Groups API, which provides the device-side \emph{grid.sync()} call.
We utilize signaling operations provided by the NVSHMEM API \cite{Nvshmem_Docs} for inter-GPU synchronization. Pairs of flags are allocated in the symmetric heap for top and bottom neighbors - four in total for each processing element - using \textit{nvshmem\_malloc()}. The flags are waited on using \textit{nvshmem\_signal\_wait\_until()} in boundary thread blocks before computation begins to ensure neighboring devices have committed values of the previous time step. The signaling flow is essentially a semaphore wherein neighboring devices signal the availability of halos of a given time step by setting the corresponding flag to the value of the next iteration while waiting is done by comparing the flag to the current iteration. The signaling operation uses the composite \textit{nvshmemx\_putmem\_signal\_nbi\_block()} call that performs data movement and subsequently updates a given flag on completion. This API is also used to write to halo regions residing in neighboring devices' memories.
Synchronizing local concurrent kernels is done by busy waiting on a local device memory flag. 

\subsubsection{Thread Block Specialization}

We specialize thread\break blocks to carry out inner domain computation and boundary computation / communication. In order to balance the two phases of execution, we adjust the number of thread blocks to reserve for boundary computation with the domain size. We use the following formula to determine work allocation.
\vspace{-1mm}
\begin{equation*} \label{boundaryTBC_eq}
    boundary\_TB\_num = \frac{TB\_total * boundary\_size}{inner\_size + 2 * boundary\_size}
\end{equation*}
\begin{equation*} \label{innerTBC_eq}
    inner\_TB\_num = TB\_total - 2 * boundary\_TB\_num
\end{equation*}
where $TB\_total$ is the total number of thread blocks available in the device for the given thread count.

Splitting the thread blocks proportionally to the amount of work is necessary for smaller domains to achieve proper overlap, as they are susceptible to being bound by the boundary region computation and communication time otherwise.

\subsubsection{PERKS Integration}
We can apply our communication scheme to existing single-GPU kernels by swapping them into our inner-computation kernel. We extend a single-GPU stencil kernel provided by PERKS \cite{perks}, as discussed in Section \ref{sec:designs:stencil}, as it performs explicit caching. 

We restrict the PERKS kernel to the inner domain while keeping it oblivious to the multi-GPU nature of execution and treating it mostly as a black box.
The PERKS kernel is minimally modified to synchronize with the communication stream at the end of each iteration, and the domain resides in memory buffers shared by both. Since both kernels access the same buffers in device memory, we need to make the PERKS process a disjoint portion without memory conflicts, which is done by making it treat the boundary layers as immutable halos. As PERKS excludes read-only halos in its caching strategy, the kernel freely observes memory writes from the boundary stream by reading from the global memory in every iteration. 

\subsection{Conjugate Gradient}

CUDA Samples implements a persistent kernel multi-GPU CG solver with Unified Memory \cite{cuda-samples-cg}. We use this sample as the starting point of our implementation but modify it heavily by opting to use NVSHMEM for communication.

To partition the domain, we equally divide the vectors across the GPUs and allocate the chunks as NVSHMEM symmetric objects. We note that NVSHMEM requires all symmetric object allocations to be the same size. To handle the case when the number of rows is not exactly divisible by the number of GPUs, we allocate a few redundant rows which are ignored in all computational kernels. For the sake of simplicity, we elect to keep a copy of the matrix on each GPU and partition it by splitting it into logical chunks.

For the CPU-free versions, we launch the persistent kernel using \textit{nvshmemx\_collective\_launch}, the cooperative launch utility function provided by NVSHMEM. Launching with this routine is required when the kernels use NVSHMEM collective or synchronization APIs, which are extensively employed in our code.

\subsubsection{Synchronization} Same as in stencils, we use the Cooperative Group API's grid.sync() function to synchronize all threads within the device. To synchronize across devices, we utilize NVSHMEM's \textit{nvshmem\_barrier\_all()} function. For the CPU-controlled baselines, we use the host-side stream-based API equivalent \textit{ nvshmemx\_barrier\_all\_on\_stream()}.

\subsubsection{Communication} All communication uses NVSHMEM calls. To get the elements on neighbor GPU's vectors for SpMV, both CPU-controlled and CPU-free versions use the \textit{nvshmem\_double\_g()} device-initiated call, a blocking call that fetches a single double from a neighbor GPU. For reducing across the dot products, the CPU-free versions use the \textit{nvshmem\_sum\_reduce\_block()}. We also experimented with the warp and thread-level reduction variants but found them to perform worse than the block-level version. 

The CPU-controlled baselines use the host-side stream-based equivalent \textit{nvshmemx\_sum\_reduce\_on\_stream()}. 
We use block-level APIs for CPU-free versions because that is an explicit advantage of using GPU-side NVSHMEM calls, as block-level APIs have no host-side equivalents. We also note that using NVSHMEM collective APIs on the device requires a \textit{cooperative kernel launch}, making it, in essence, a persistent kernel. 

\subsection{Limitations}
The CPU-free execution model suffers from occupancy limitations set by the Cooperative Groups API due to its persistent nature.
Device-wide barriers are only supported in cooperative launches that do not oversubscribe the number of thread blocks, meaning it is impossible to request more blocks than the device can run concurrently.

For larger domains, such distribution of work is instead delegated to the programmer (or software), who is required to partition the domain into successive tiles and iterate over them. 
This limitation is imposed to guarantee the co-residency of the thread blocks, as synchronizing across the grid would otherwise be impossible.
We apply such successive tiling in our implementation to assign workloads to threads, which can introduce inefficiencies compared to the hardware scheduler in large domains.

%% file: sections/results.tex
\begin{figure*}[!htbp]
    \centering
    \includegraphics[width=\linewidth]{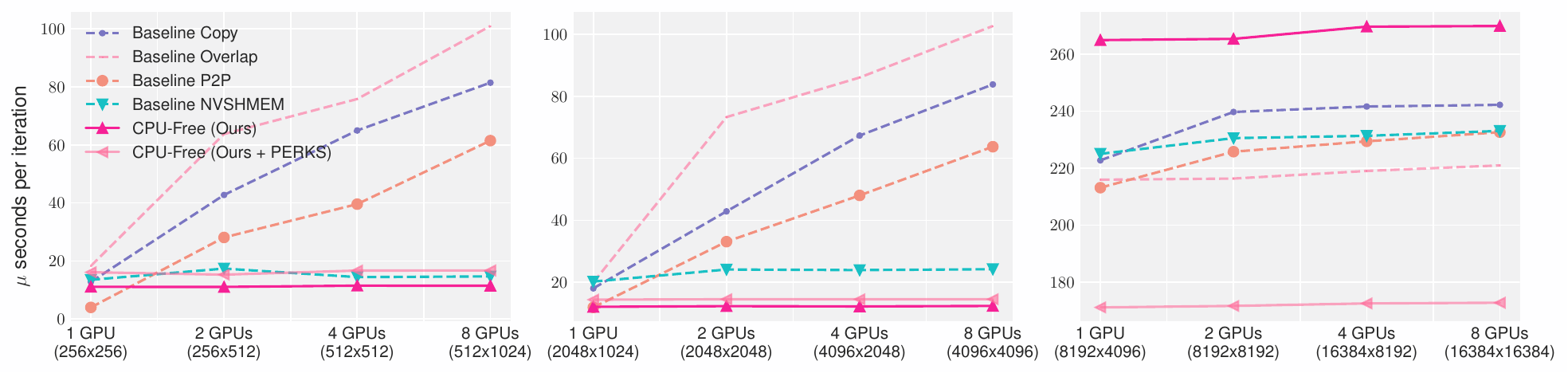}
    
    \caption{Weak scaling of 2D Jacobi stencil method with small to large domain sizes on up to 8 NVIDIA A100 GPUs}
    \label{fig:2D-Weak-Scaling-all}
\end{figure*}

\begin{figure*}[!htbp]
    \centering
    \includegraphics[width=\linewidth]{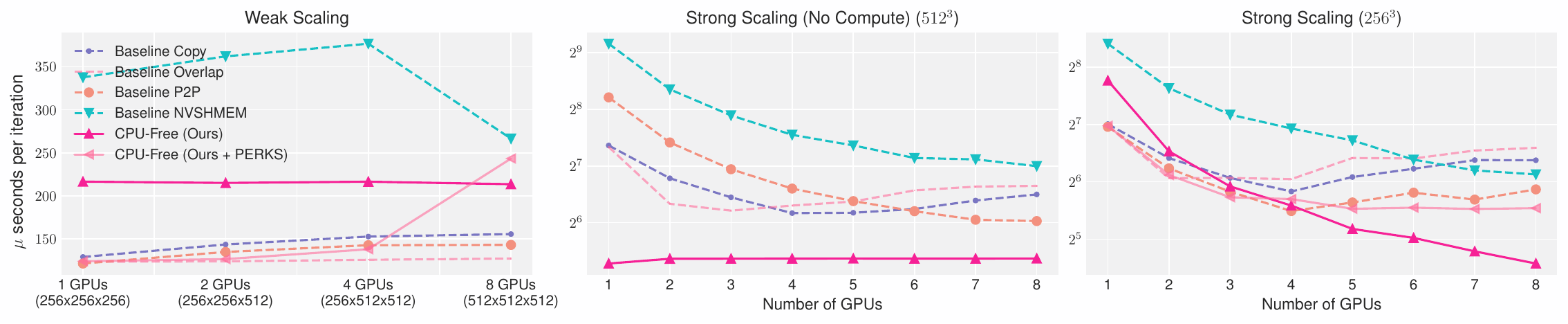}
    
    \caption{3D Jacobi stencil weak and strong scaling experiments on up to 8 NVIDIA A100 GPUs}
    \label{fig:3D-Scaling-all}
\end{figure*}

\section{Evaluation}
\label{sec:eval}

This section presents the performance scaling of CPU-free execution over CPU-controlled execution for the 2D/3D Jacobi stencil and Conjugate Gradient methods.
The experiments presented here were conducted on NVIDIA HGX machines with 8 NVIDIA A100s connected through NVLink with CUDA toolkit version 11.8 and driver version 495.29.05. The NVSHMEM library version is 2.7.0 with OpenMPI 4.1.4.
We repeat each experiment 5 times and report the minimum. 

\vspace{-2mm}
\subsection{Stencil Benchmarks}
\subsubsection{Experiment Setup and Code Variants}

We categorize the experiments into three groups of domain sizes: small, medium, and large.
The domain sizes are categorized as such based on device saturation: a small domain has fewer elements than needed to keep the entirety of the device busy, while medium has sufficient elements to utilize all thread blocks, and large domains over-saturate the device. For a 2D5pt Stencil on an NVIDIA A100 with 108 SMs, we select our domains as $256^2$, $2048^2$, and $8192^2$, respectively.

The baselines are taken from NVIDIA's multi-gpu programming models  repository \cite{nvidia-multi-gpu-programming-models}, which contains Jacobi kernels with different communication schemes. We compare our implementations to the four best-performing versions:

\begin{itemize}[align=left,leftmargin=*,widest={10}]
    \item \textbf{Baseline Copy}: Standard CPU-controlled implementation with no explicit boundary overlap. This version only overlaps memory transfers with kernel execution using host-side asynchronous cudaMemcpy calls.
    \item \textbf{Baseline Overlap}: Same as above, but computes boundary rows in separate streams independently of the inner domain for explicit overlap, and synchronizes using host-side events. The explicit overlap this version performs is identical to our implementation.
    \item \textbf{Baseline P2P}: Communication is done through device-side direct load and stores in peer-to-peer memory. Although the communication is GPU-initiated in this version, synchronization is handled by the host.
    \item \textbf{Baseline NVSHMEM}: Uses device-side NVSHMEM calls for communication. This version utilizes the same family of NVSHMEM memory operations that we use in our CPU-Free implementation, except in CPU-controlled discrete kernels. It additionally uses a dedicated kernel to synchronize neighboring GPUs to avoid redundantly synchronizing all processing elements. Both kernels are launched by the CPU in every time step.
    \item \textbf{CPU-Free}: Our implementation as described in Section \ref{sec:impl:stencil}.
\end{itemize}

We additionally measure multi-GPU PERKS performance with our communication scheme, as it tiles the compute kernel to overcome the limitations of the Cooperative Groups API.  

Lastly, we implement a 3D7pt stencil partitioned across the $z$ axis, and adapt existing 2D baselines to 3D.

\subsubsection{Scaling Experiments}

Figure \ref{fig:2D-Weak-Scaling-all} shows weak-scaling\break measurements of per-iteration time in the aforementioned domain sizes.

We vary the domain across all axes in alternating order as we double the number of devices for our weak scaling study.
At 8 GPUs, we achieve speedups of $41.6$\% and $48.2$\% in small and medium domains respectively over the best-performing baseline (Baseline NVSHMEM), and $96.2$\% and $95.7$\% speedup over the fully CPU-Controlled Baseline Copy Overlap.

It should be noted that the performance degradation we experience in the largest domains compared to baseline is due to subpar tiling in the computational kernels. As per the limitations set by the cooperative groups, kernels can only be launched in configurations that guarantee the co-residency of blocks, meaning, in our case, we only allocate one block of 1024 threads on each SM. In order to process the entire domain, we tile threads in software, which causes  inefficiencies in large domains. We display multi-GPU PERKS results as an alternative, as their computational kernel provides better tiling. PERKS kernel with our communication scheme achieves good weak scaling within a $9$\% dropoff at 8 GPUs, and $18.8$\% speedup over the baseline on the largest domain. 

Figure \ref{fig:3D-Scaling-all} shows 3D Jacobi performance.

Though the weak scaling displays lower overall performance due to the large domain, we measure better no-compute time, which is shown in the same figure in the middle at its largest domain.
Figure \ref{fig:3D-Scaling-all}, also conducts strong scaling experiments on a constant large 3D domain to demonstrate synchronization overheads and the overlap amount as the number of devices increases. We notice that the CPU-Free version stays largely flat, while the CPU-controlled baselines degrade. 

With a small number of GPUs, each GPU has a large domain that is limited by computation, not communication.
However, as the GPU count increases, communication and overheads become dominant, and CPU-free clearly exhibits its communication advantages, as seen in Strong Scaling (No Compute) experiments.

In summary, CPU-free may not always be optimal, but it has proven to be highly effective for small to medium domains when CPU-induced latencies consume a substantial portion of the runtime and communication/computation can be sufficiently overlapped.
Our approach excels in strong scaling scenarios where the overhead ratio increases with GPU count.
Conversely, traditional CPU-controlled implementations fail to achieve overlap, causing communication and computation to be serialized in small-medium domains.

\subsection{Conjugate Gradient}
\subsubsection{Experiment Setup and Code Variants}
From the SuiteSparse Matrix Collection, we select several symmetric positive definite (SPD) matrices that can converge in a CG solver \cite{suitesparse}. 

We select a wide range of matrices to showcase as broad application scaling behavior as possible. The matrices are summarized in Table~\ref{table:matrices}, sorted by the number of non-zeros.

We compare our CPU-free versions against CPU-controlled baselines as explained below: 

\begin{itemize}[align=left,leftmargin=*,widest={10}]
     \item \textbf{CPU-Controlled Standard CG}: Standard CG with discrete kernels and CPU-initiated communication.
    \item \textbf{CPU-Controlled Pipelined CG}: Pipelined CG with discrete kernels and CPU-initiated communication. Uses concurrent GPU streams for overlap. One stream computes SpMV while the communication stream performs the dot reductions. The communication stream is launch\-ed with a higher priority so that it is always scheduled first.
    \item \textbf{CPU-Free Standard CG}: Standard CG with the CPU-free model. 
    \item \textbf{CPU-Free Pipelined CG}: Pipelined CG with the CPU-free model. Uses TB specialization for overlap as described in \textit{Section \ref{subsec:conjugate-gradient-use-case}}.

\end{itemize}

\begin{figure*}
    \centering
    \includegraphics[width=\linewidth]{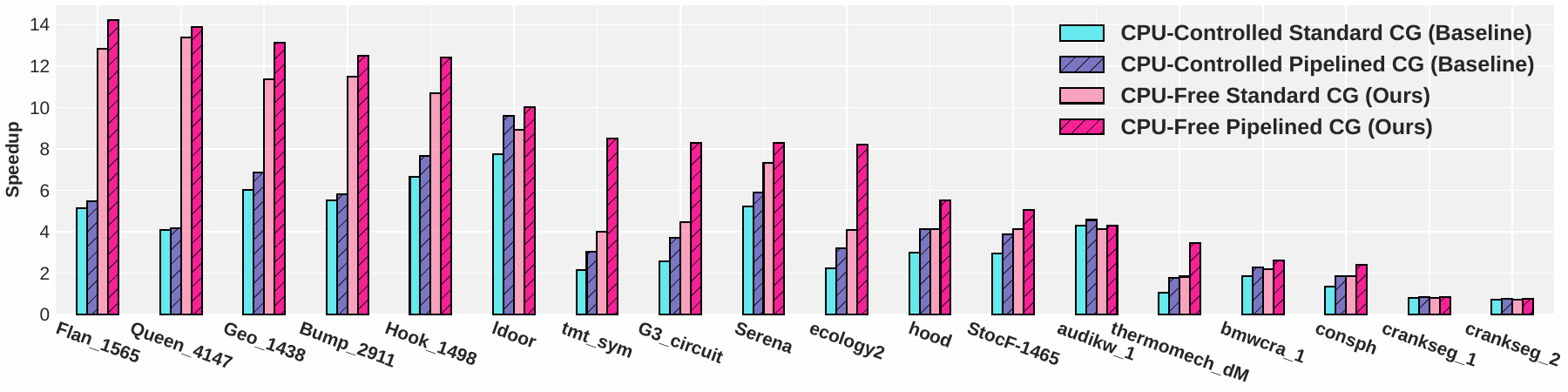}
    \caption{Speedup over Single GPU Standard CG of various sparse matrices on 8 NVIDIA A100}
    \label{fig:CG-speedups}
\end{figure*}

\begin{figure*}[!htbp]
    \centering
    \includegraphics[width=\linewidth]{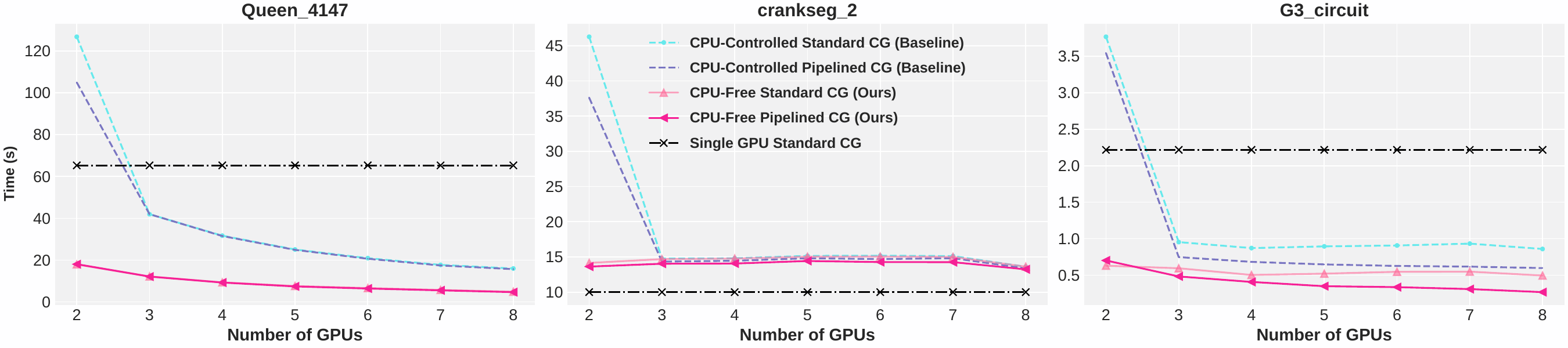}
    \caption{Strong scaling study of CG for selected matrices with large, medium, and small sizes}
    \label{fig:cg-strong-scaling}
\end{figure*}

Additionally, we implement \textbf{Single GPU Standard}, a pure single GPU version of Standard CG with no communication. We use this reference version for reporting speedup. We note that we treat the CPU-controlled baselines fairly to the best of our ability. All versions share the same control flow, placement of within-device and across-device barriers, computational routines with a thread block size of $1024$, and communication calls. The only difference between the versions is whether they use the host or device-side APIs. 

While, in practice, the CG algorithm is often coupled with a preconditioner for faster convergence \cite{preconditioned-cg}, we do not apply it as it is orthogonal to our focus on communication. Additionally, we run both \textit{Standard} and \textit{Pipelined} CG algorithms for a fixed number of iterations (5000) and not necessarily to convergence. 

Even though these algorithms have different convergence properties, we consider the numerical stability of either method to be outside the scope of this work.

\input{listings/matrices.tex}

\subsubsection{Scaling Study} Figure \ref{fig:CG-speedups} shows the speedups achieved by all versions over the {\em Single GPU Standard} variant. Regardless of the CG variant, CPU-free versions outperform CPU-controlled baselines on almost all matrices, but the speedup varies across matrices. As also shown in Table \ref{table:matrices},  CPU-Free Pipelined CG achieves 1.63x geometric mean speedup over CPU-Controlled Pipelined CG  while CPU-Free Standard CG gets 1.54x geometric mean speedup over CPU-controlled Standard CG. 

For most of the largest matrices (i.e. \texttt{Queen\char`_4147}, \texttt{Bump\char`_291}) 

we observe the significant speedup of our versions over the CPU-controlled ones. 

For these matrices, SpMV takes up the most time, and there is not enough communication to achieve good overlap. We can observe this by looking at \texttt{Queen\char`_4147} in Figure \ref{fig:cg-strong-scaling}. For both CPU-free and CPU-controlled versions, there is little difference between the Pipelined and Standard variants meaning there is almost no overlap. Another implication of this is that the speedups we observe are a direct consequence of 
moving the control path to GPU.

For other matrices (\texttt{audikw\char`_1},  \texttt{crankseg\char`_2}, \texttt{crankseg\char`_1}, \texttt{ldoor}, \texttt{bmwcra\char`_1}), we observed comparatively subpar speed\-up compared to the larger matrices. While these matrices have a smaller number of non-zeros, we observe that they are irregular. When these matrices are split among GPUs, there is a significant load imbalance; some GPUs spend more time in SpMV than others. The implication is that all versions are constrained by single-GPU performance. 

We predict that these matrices could benefit from matrix reordering. 

On the other end, we observe significant speedups for smaller matrices (G3\_circuit, tmt\_sym, ecology2, thermomech\_dM). We note that these matrices have the lowest sparsity 
(nnz/row); thus, communication takes up a more significant portion of the runtime. In this case, good communication-computation overlap can be achieved, which we observe by comparing the speedups for Pipelined and Standard variants.

%% file: listings/matrices.tex
\begin{table}[!tb]
\caption{SuiteSparse  matrices used in CG evaluation}
\centering {\footnotesize
\begin{tabular}{ccccc} \toprule
    {\textbf{Matrix}} & {\textbf{Rows}} & {\textbf{NNZ}} & {\textbf{NNZ/Rows}} & {\textbf{Speedup}}  \\ \midrule
    \texttt{Queen\_4147}    & 4,147,110 &  329,499,284 & 79.45 & 3.33x \\
    \texttt{Bump\_2911}     & 2,911,419 & 127,729,899 & 43.87 & 2.15x \\
    \texttt{Flan\_1565}     & 1,564,794 & 117,406,044 & 75.03 & 2.59x \\
    \texttt{audikw\_1}      & 943,695   & 77,651,847  & 82.28 & 0.94x \\ 
    \texttt{Serena}         & 1,391,349 & 64,531,701  & 46.38 & 1.4x \\ 
    \texttt{Geo\_1438}      & 1,437,960 & 63,156,690  & 43.92 & 1.91x \\ \midrule
    \texttt{Hook\_1498}     & 1,498,023 & 60,917,445  & 40.67 & 1.62x \\ 
    \texttt{ldoor}          & 952,203   & 46,522,475  & 48.86 & 1.04x \\ 
    \texttt{StocF-1465}     & 1,465,137 & 21,005,389  & 14.34 & 1.3x \\
    \texttt{crankseg\_2}    & 63,838    & 14,148,858  & 221.64 & 1.0x \\ 
    \texttt{hood}           & 220,542   & 10,768,436   & 48.83 & 1.34x \\ 
    \texttt{bmwcra\_1}      & 148,770   & 10,644,002  & 71.55 & 1.15x \\ \midrule
    \texttt{crankseg\_1}    & 52,804    & 10,614,210  & 201.01 & 1.0x \\ 
    \texttt{G3\_circuit}    & 1,585,478 & 7,660,826   & 4.83 & 2.23x \\
    \texttt{consph}         & 83,334    & 6,010,480   & 72.13 & 1.3x \\ 
    \texttt{tmt\_sym}       & 726,713   & 5,080,961   & 6.99  & 2.8x \\
    \texttt{ecology2}       & 999,999   & 4,995,991   & 5.00 & 2.56x \\
    \texttt{thermomech\_dM} & 204,316   & 1,423,116   & 6.97 & 1.94x \\  \midrule
    \multicolumn{4}{c}{CPU-Free Pipelined Speedup over CPU-Controlled (geo. mean)} & 1.63x   \\ \bottomrule 
    
\end{tabular}
}
\label{table:matrices}
\vspace{-4mm}
\end{table}

%% file: sections/related-work.tex
\section{Related Work}
Previous work has shown the benefits of reducing CPU involvement in GPU execution, both along the directions of persistent kernels and GPU-initiated communication. The work by Kshitij et al. \cite{persistent-threads} was the first to summarize the persistent kernel concept and discuss how persistent kernels help better load balancing and kernel fusion. Uberkernels that result from the fusion of multiple stages of a dynamic application are proposed in WhippleTree, mostly focusing on graphics workloads \cite{whippletree}. 

More recently, work by Zhang et al. \cite{perks} demonstrates significant speedup over state-of-the-art single GPU stencil and CG baselines by using shared memory and registers to avoid global memory accesses. 

Chu et al. \cite{gpu-key-value} apply persistent kernels to GPU-based key-value stores to reduce host-induced kernel launch and memory copy overheads leading to better scaling and communication-computation overlap. 

Several works have provided the building blocks toward the ultimate goal of freeing the GPU by reducing the involvement of the CPU in the control path. Early works experimented with running the entire network stack on the device but, at the time, suffered from subpar performance and correctness issues due to GPU-NIC interaction \cite{GPUrdma, infiniband-verbs-on-GPU}.  Other works added support for GPU networking through CPU helper threads \cite{dCUDA, GPUNet}. Agostini et al. \cite{Agostini2018GPUDirectAsync} introduced GPUDirect Async, which optimized the control path between GPU and NIC by allowing the GPU to trigger and sync CPU-enqueued network transfers.  Later, \cite{GPU-Triggered-Networking} proposed a NIC hardware mechanism that allowed GPUs to trigger CPU-registered network operations on the NIC from within the kernel, bypassing the CPU, and showed promising results on a simulation. GIO examined the GPU's relaxed memory model for AMD GPUs and fixed the correctness issues stemming from GPU-NIC interaction \cite{GIO}. Work on NVSHMEM has also optimized for both the data and control paths by providing APIs for fine-grained GPU-to-GPU data movement from within CUDA kernels \cite{nvshmem-1, nvshmem-2, nvshmem-3, nvshmem-4}. While previously NVSHMEM launched CPU proxy threads when communicating over InfiniBand, as of NVSHMEM 2.6.0, the CPU proxy thread can be bypassed, allowing kernel-initiated communication to be issued directly to the NIC \cite{nvshem-release-notes}. 

Other works have used both persistent kernels and GPU-controlled communication to introduce execution models and runtime systems for different applications. Work by Belviranli et al. \cite{juggler} uses a persistent kernel to implement a task-based execution model that treats thread blocks as standalone execution units - a more general form of our block specialization scheme. Chen et al. \cite{atos, atos-single} present a task-scheduling framework for irregular applications using both persistent and discrete kernels.